\documentclass[10pt, conference, letterpaper]{IEEEtran}
\IEEEoverridecommandlockouts
\usepackage{cite}
\usepackage{amsmath,amssymb,amsfonts}
\usepackage{algorithmic}
\usepackage{comment}
\usepackage{enumitem}
\usepackage{graphicx}
\usepackage{booktabs} 
\usepackage{textcomp}
\usepackage{xcolor}
\def\BibTeX{{\rm B\kern-.05em{\sc i\kern-.025em b}\kern-.08em
    T\kern-.1667em\lower.7ex\hbox{E}\kern-.125emX}}
    
\usepackage{tikz}
\usepackage{textcomp}
\usepackage{hyperref}
\usepackage{lipsum}
\newcommand\copyrighttext{%
  \footnotesize \textcopyright 20xx IEEE. Personal use of this material is permitted.
  Permission from IEEE must be obtained for all other uses, in any current or future 
  media, including reprinting/republishing this material for advertising or promotional 
  purposes, creating new collective works, for resale or redistribution to servers or 
  lists, or reuse of any copyrighted component of this work in other works. 
  }
\newcommand\copyrightnotice{%
\begin{tikzpicture}[remember picture,overlay]
\node[anchor=south,yshift=10pt] at (current page.south) {\fbox{\parbox{\dimexpr\textwidth-\fboxsep-\fboxrule\relax}{\copyrighttext}}};
\end{tikzpicture}%
}

\begin{document}

\title{GDPR: When the Right to Access Personal Data Becomes a Threat}

\author{\IEEEauthorblockN{Luca Bufalieri, Massimo La Morgia, Alessandro Mei, Julinda Stefa}
\IEEEauthorblockA{Department of Computer Science, Sapienza University of Rome, Italy\\\\
Email: bufalieri.1430586@studenti.uniroma1.it, \{lamorgia, mei stefa\}@di.uniroma1.it}}
\maketitle
\copyrightnotice

\begin{abstract}
After one year since the entry into force of the GDPR, all web sites and data controllers have updated their procedure to store users' data. The GDPR does not only cover how and what data should be saved by the service providers, but it also guarantees an easy way to know what data are collected and the freedom to export them.

In this paper, we carry out a comprehensive study on the right to access data provided by Article 15 of the GDPR. We examined more than 300 data controllers, performing for each of them a request to access personal data. We found that almost each data controller has a slightly different procedure to fulfill the request and several ways to provide data back to the user, from a structured file like CSV to a screenshot of the monitor. We measure the time needed to complete the access data request and the completeness of the information provided. After this phase of data gathering, we analyze the authentication process followed by the data controllers to establish the identity of the requester. We find that 50.4\% of the data controllers that handled the request, even if they store the data in compliance with the GDPR, have flaws in the procedure of identifying the users or in the phase of sending the data, exposing the users to new threats. With the undesired and surprising result that the GDPR, in its present deployment, has actually decreased the privacy of the users of web services.
\end{abstract}


\section{Introduction}

The General Data Protection Regulation (GDPR)~\cite{gdprlaw} is a regulation of the European law on data protection and privacy. The regulation protects personal data of natural persons and lays down the rules relating to the free movement of personal data. The GDPR entered into force in May 2018, and all the entities that control data and are located or offer services in the European Economic Area (EEA) must comply with it. The law applies to entities that monitor behaviors that take place in the European Union, even if they do not provide any direct service.
Article 15 of the law establishes one of the most fundamental rights in the Internet: The users have the right to request to web sites a copy of all the personal data that they have about them. The goal of this right is to return control and awareness to the users of the personal data they share, consciously or not.

In this frame, we carry out a broad world-scale investigation on the actual deployment of the GDPR. We perform a step-by-step analysis of all the phases needed to accomplish a subject access request (SAR)---the action of request the personal data to a data controller. In our study, we target 334 of the most popular web sites according to the Alexa ranking. For the best of our knowledge, we are the first to conduct a comprehensive study on this topic with a world distribution of web sites, so our finding are also useful to refine previous works that took into account only one phase of the SAR~\cite{wang2018next}, or used less rigorous methodologies to select the organizations~\cite{pavur2019}, or could be biased by the small set of data controllers put under the lens~\cite{urban2019study}.

We find that 19.6\% of privacy policy pages are not compliant with the actual regulation. Then, we inquiry all the targeted web sites requiring our personal data. We study how the collectors identify the requester, we collect the response, and monitor the response time. In the end, we obtain our personal data from almost 65\% of the targeted web sites, with a average time to fulfill the request of 16.4 days. 
Lastly, we checked the procedures used by the data controllers to fulfill the request. In this phase, we find several flaws that affect more than 32\% of targeted data controller, and that could transform a fundamental right into a new and unpleasant threat.

This paper makes the following contributions:
\begin{itemize}
    \item \textbf{World-wide snapshot:} We makes a world-wide snapshot of the actual deployment of the GDPR. We report on the aggregated metrics about the privacy policy compliance, the methodology to request the data, the identification process, the response time, and the response format.
    \item \textbf{Response analysis:} We analyze the obtained responses and the data collected by the web sites. Here we find a clear lack of information on the data returned by the data controllers. For instance, the same data controller return a different amount of information depending of the type of request we made (Right of Data Access or Right of Data Portability). E-shops that do not return information at all about the visited pages, then use these information in re-marketing campaigns. 
    \item \textbf{Vulnerability analysis:} We perform a vulnerability analysis on the methodology adopted by the data controllers to transmit the data to the requester. During our collection phase we find that also the collectors that perform very well in the identification phase jeopardize the data in the transmission phase, like sending the data as plain text or not using secure protocols to send the email.

    \item \textbf{Identification robustness:}  We investigate on the two most particular cases we found in the identification process: The data controllers that don't seem to care about the identity of the requester and those that want an ID document to correctly identify the requester. In the first case, we discovered that 8.9\% of data controllers actually disclose the personal data of the users regardless the real identity of the requester. While, in the second case, we bear out that the data controllers do not have proper tools to verify the originality of the ID (i.e. if the ID has been tampered).
\end{itemize}

\section{Research design}

\subsection{Selecting the organization to test}
To perform our analysis, we gathered the top 50 web sites\footnote{On May 2019} from each of the 12 categories of the Alexa ranking.
From the list, we removed duplicates, e.g.\ the same web site can appear under multiple categories, or the same company can own multiple web sites.
Then, we manually visited each web site. We discarded those who do not provide pages in languages spoken in the EU because we expect that the web sites that are not based in the European Economic Area and offer services to European people have at least a version of the web site in one of the languages spoken in Europe. We removed from the list all the web sites that do not provide a registration procedure. Indeed, we focus on websites that store personal data linked to identified users.
In the end of the process we selected 334 web sites of the following categories: Adult (7.5\%), Art (5.1\%), Business (7.8\%), Computer (9.6\%), Games (8.4\%), Health (1.8\%), Kids \& Teen (2.8\%), News (7.2\%), Recreation (12.9\%),  Reference (5.7\%), Science (3.9\%), Shopping (15.6\%), Society (6\%), and Sports (5.4\%).
Since the regulation does not apply only to web sites, we inserted in the list other data collectors---2 banks, 3 transport companies, and 2 mobile network operators. For the latter categories of data providers, it is neither easy to create new accounts nor generate meaningful data. So, we select providers for whom we owned a long-time active account. Since no third-party trackers or advertisement networks were in the list, we manually added to the list 5 of them after the polishing phase. 
To select the trackers, we randomly picked 5 active trackers on our browser from the list reported by the online tool \textit{youronlinechoices}\footnote{http://www.youronlinechoices.com}. This tool is provided by the EDAA---European Interactive Digital Advertising Alliance, a European industry coalition of advertising agencies. 

\subsection{The sign-up phase}
As said, we want to be sure that the data controllers have personal data to return on our request. So, we proceed to create a new account for each web site. Then, we manually used the web site for a while, performing actions that the web site has to log: adding items to favorites, playing videos, making queries, interacting with the email they send, and so on. During the registration phase, we discarded 5 web sites because they did not allow registration. Even though these web sites were accessible from Europe, we found out that the registration form was disabled for European users (indeed, we checked that these websites do accept registrations from outside the EU).
At the end of the process we have a list of 341 data collectors to request personal data from.


\subsection{Leading phases of a subject access request}
Once we created and initialized all the accounts, our next step is to request the personal data.
During our experiment, we focused on the most relevant aspects of the process: privacy policy compliance, the request methodology, the response time, the response format, and the completeness of the information.  In the following sections, we explain why we focused on these aspects and how we evaluated the responses.

\subsubsection{Privacy policy compliance}
The GDPR states that the data collectors must inform the users about the rights of requesting a copy of, updating, or deleting their personal data owned by the data controllers. Moreover, they also must provide the contact details of the data protection officer (DPO). 
The privacy policy page is the place where the users shall be informed about their rights, so we check if these pages comply with the GDPR.



\subsubsection{Request methodology}
Data collectors shall facilitate the exercise of subject data rights (GDPR Article 12 sec. 2). We focus on the request methodology because we believe that it represents the primary access to obtain personal data. A web site that accepts requests only via postal mail or an international phone call, phone call can discourage the user from exercising his rights. Conversely, one that offers an online form facilitates the accessibility to this information.

\subsubsection{Identification}
Since we deal with personal data, it is crucial that only the owner can access the information collected by the data controller. Hence, we are interested in understanding how the controllers ensure that data is provided only to the right person.

\subsubsection{Response format}
\label{sec:resp_format}
We take into account the type of data format used by the web sites to provide the data. 
The GDPR defines two rights that allow the users to access their personal data.
The first one is the right to access. In this case, to the best of our knowledge, there are no strict constraints about the response data format (Article 15). The second is the right of data portability: the data should be provided in a commonly used and machine-readable format (Article 20).


\subsubsection{Response time}
Once we received the response, we computed the amount of time needed by the web sites to send back the personal data. Indeed, the data controllers have 1 month to process the request and provide the data. If necessary, the controller can extend that period for additional two months (GDPR Article 12 sec. 3), but it must communicate the delay to the user before the end of the first month. Our goal is to check if the controllers satisfy the timing constraint of the GDPR. Moreover, we want to understand what is the average time needed by the controllers to fulfill the request.


\subsubsection{Information obtained}
As the final step, we report on the information retrieved by the data controllers, and inconsistencies we found.


\section{Results}

\subsection{Privacy policy compliance}
\label{sec:p&p}
To begin our investigation, we want to understand if the web sites transposed the GDPR law and updated their privacy policy accordingly. 
We consider a web site compliant with the GDPR if it specifies the user's rights and a contact point for privacy information on its privacy policy page, or in any of its pages reachable by a Google query.
To compose the query, we use the name of the web site plus one or a combination of the following key terms: "GDPR", "users right", "data access", "Subject Access Request".

Among all the web sites targeted, we found that 6 out of 341 of them mention neither the user's right of data access in their privacy policies nor indicate a contact point.
53 (15.54\%) web sites do not mention the users' rights, while 3 (0.87\%) web sites do not have any contact point for privacy information. Finally, the privacy policy pages of 3 web sites (0.87\%) do not work.
Consequently, we had to remove 9 web sites since we had no way to request the data, leaving in our list 332 web sites.

Analyzing the results, we found that 4 out of 6 data controllers, that do not report both the rights and the contact point, are forums or small services related to the video games world. These findings make us believe that complying with the GDPR law could be a problem, especially for small sites and services that are on the low positions of the Alexa global ranking. However, they can be very popular for a specific niche of users.
Among the 53 data controllers that do not mention users' rights, most of them are located outside the EEA. In their privacy policy, we find 5 of them that explicitly discourage European people from using their platform, but at the same time, they allow them to sign up. While 22 out of 53 web sites have privacy policy pages that has not been updated in the last two years, according to the last update date they report on the web page.


\subsection{Request Methodology}
Every web site implements its procedure to start a subject access request. Usually, we found the procedure to exercise the right to access the data described in the privacy policy page of the data controller. If no information is present on this page, we contact the DPO via email. We found 4 main ways to ask the data. In particular, for 219 out of 332 (65.9\%) data controllers it is enough to send an email, while for 11 (3.3\%) data controllers the user also need to compile a form and send it by email. For 96 (28.9\%) data controllers it is possible to request the personal data through a form on the web site, we also consider in this category the 17 web sites that expose a button to immediately download the personal data. Finally, For 6 (1.8\%) data controllers, the only way to perform the request is by standard mail or by an phone call outside Europe. We leave out from our experiment this last set of data controllers and we perform the SAR to the remaining 326.  
\subsection{Identification}
Before transmitting the data, the DPO is supposed to verify the identity of the requester. As required by law, the DPO in this phase can ask for additional information if he has reasonable doubts concerning the identity of the requester. We found that web sites that implement a form to perform the request do not require a further identification if the form is reachable only by a signed-in user. 2 most scrupulous web sites request to re-authenticate to finalize the request. Instead, other data controllers rely on one of the following mechanisms or a combination of them: identification document, sworn declaration, phone call, questions about user personal data (e.g.\ date of birth, username, address), cookie ID, address of the email sender, questions about data that can be retrieved only from inside the account. Table~\ref{tab:group_users} shows the number of data controllers that use a specific mechanism to identify the user, associated with each way to ask the data. Between parenthesis are reported the number of data controllers that use a combination of mechanisms. 

\begin{table*}
	\centering
	\small
    \caption{%
        User identification mechanism by type of request.
    }\label{tab:group_users}
    \begin{tabular}{l c c c c c c c c c }
        \toprule
        Request  & No Identification & ID & Log-in & Confirmation & Questions & Cookie  & Phone call & Sworn Declaration\\
        \midrule
        Email &  51& 22 (3) &  10 & 10 & 15 (3) & 2 & 3 & 2 \\

	    Form & 7 & 2 &  0 & 0 & 0 & 1 & 0 & 0\\

     Online Form & 0 & 9 (3) &  59  & 21 (1) & 4 (3)   & 0 & 0 & 1 (1)   \\
       
        \bottomrule
        Total & 58 & 33 &  69  & 19 & 18 & 3 & 3  & 3   \\
    \end{tabular}
    \\
\end{table*}
\subsection{Response obtained}

\begin{figure}
\includegraphics[width=0.4\textwidth]{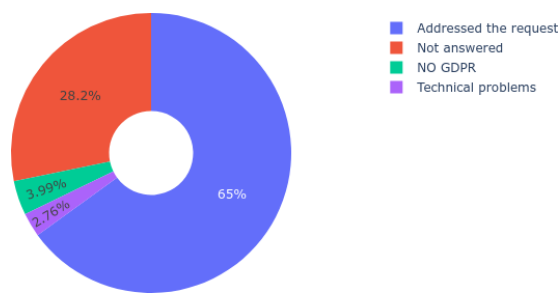}
\caption{Percentage of handled SAR request}
\label{fig:response}
\end{figure}
In Fig.~\ref{fig:response} are shown the response we obtained from the 326 data controllers. We considered 92 (28.22\%) web sites as data controllers that not handled the request, all the data controllers that did not answer to the data subject access request within 30 days from the day we forwarded our inquiry, and did not ask to extend the 30 days period.
However, several data controllers answered after the law period expired, so it is possible that, eventually, these data controllers would have answered.
For 9 (2.7\%) of data controllers we targeted in our experiment, it was not possible to deliver our requests for technical problems, such as the form did not work or the email of the DPO is out of space. Moreover, 13 (3.98\%) of controllers refused to provide us with the requested data because, in their opinion, they are not affected by the GDPR law.
Finally, for 212 out of 326 (65.03\%) of data controllers, we were able to obtain our personal data.

Table~\ref{tab:gdpr_compliance} summarizes the results obtained. In the table, we split the data controllers into two sets: the controllers that report about the GDPR rights in their privacy policy and those that do not. Indeed, as we said in Sec.~\ref{sec:p&p}, 53 data controllers did not mention the users' rights in their privacy policy. However, we performed the SAR request also to them, to understand how these data controllers react to our request. Surprisingly, we found that 17 data controllers correctly handled the request. While, as we can see from the Table~\ref{tab:gdpr_compliance} none of the data controllers that report about the GDPR rights explicitly refused to provide the data.

\begin{table}
	\centering
	\small
    \caption{%
        Response obtained by GDPR rights.
    }\label{tab:gdpr_compliance}
    \begin{tabular}{l c c}
        \toprule
        &GDPR Rights & No GDPR Rights\\
        \midrule
        Answered & 195 & 17 \\
        Did not answer & 69& 23\\
        Refused & 0 & 13 \\
        \bottomrule
        Total & 264 & 53 
    \end{tabular}
\end{table}

\subsection{Response time}
The response time of the data controllers is hugely different. It can vary from a few seconds to more than 90 days. Of course, controllers that immediately provide access to personal data are the ones that have a fully automated procedure, from the beginning of the request to the data provisioning.

\begin{figure}
\begin{center}
\includegraphics[width=0.45\textwidth]{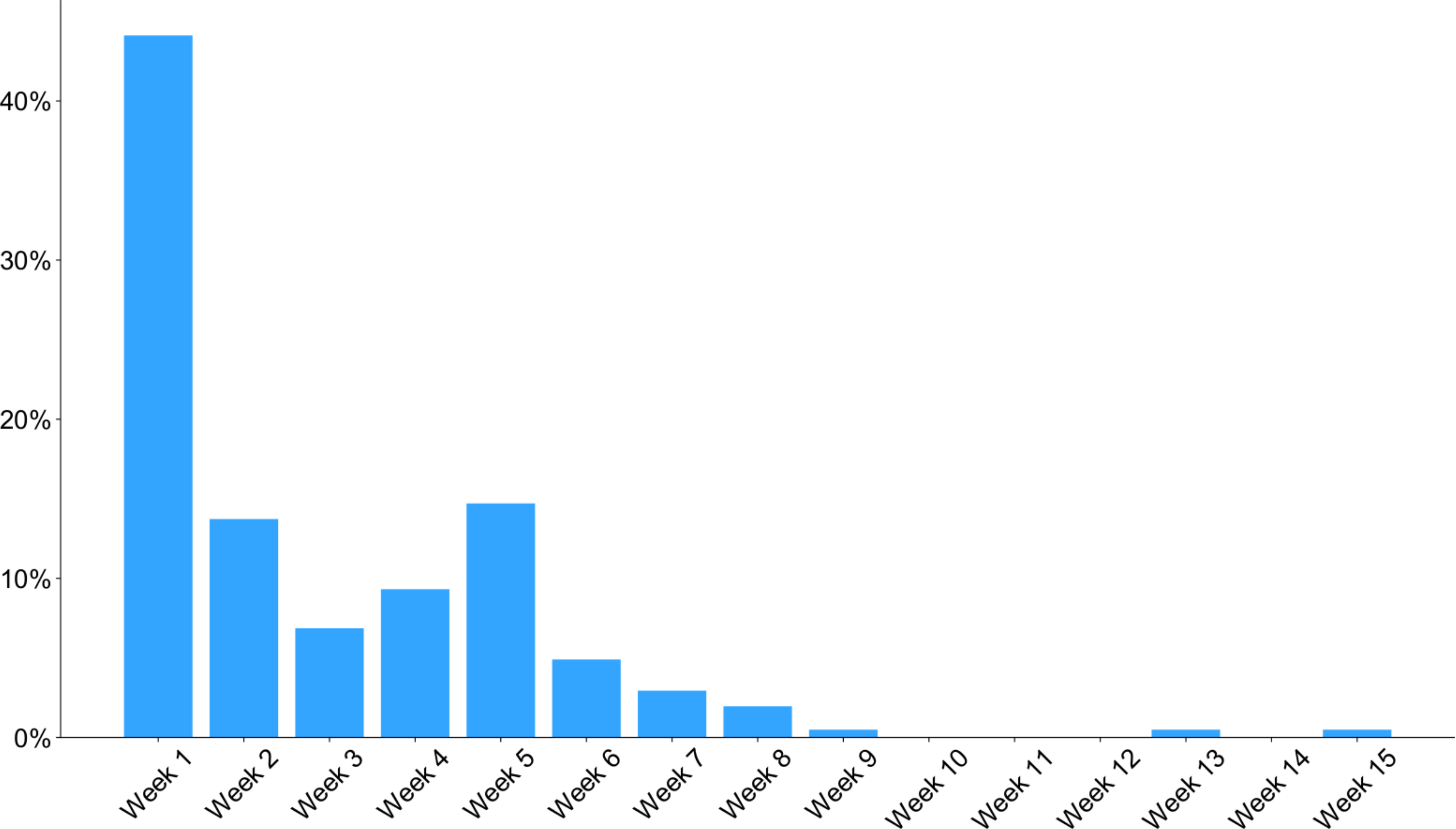}
\caption{Percentage of  responses obtained weekly.}
\label{fig:response_time}
\end{center}
\end{figure}

In Fig~\ref{fig:response_time} are shown the percentage of response obtained with a weekly granularity. 
Among the 212 controllers that handled the request, 17.45\% answered on the same day of the request, 39.62\% in the first half of the month, 26.88\% in the second half, and 13.67\% in the second month. 5 data controller needed more than two months, and 4 more than three. 
To summarize, almost all the data controllers satisfy the timing dictated by the law, with the exception of 23 data controllers. 
On average, the controllers needed 16.4 days to fulfill our requests.
Globally, we obtain that 89.15\% of controllers that handled our request comply with the GDPR. This result is in contrast with the one obtained in~\cite{urban2019study} in which they found that only the 55\% of respondent controllers handled the requests in time. This difference is probably due to the more significant amount of controllers taken into account in our study.

\subsection{Response Format}
A large number of controllers (52.7\%) answered with a structured data format---JSON, CSV, XLS or XML--- following up a data access request.
This result is surprisingly high if we compare it with the finding of Wong et al.~\cite{wong2018portable}. In their case, they obtained only 55\% of structured files after exercising the right of data portability.
So, we believe that these data controllers have a unique procedure to handle both the requests for data access and for data portability.
The rest of the controllers provided us the personal data as raw text or in a tabular form, directly typed in the email body or as a PDF attachment. 2 controllers printed our data and sent it to us by postal mail. We also collected: screenshots of the management software, scan of printed files, and HTML pages.
In the end, there are controllers that offer in addition to the possibility to download the data, also a dashboard inside the account. Here, it is possible to explore efficiently and manage the personal data collected by them in real-time.

\subsection{Information achieved}
As we can expect, all the data controllers provide us all the information that we provided in the sign-up phase. The laziest controllers, fortunately few, answered us to check ourselves the information in our profile page. However, checking the privacy policy page of these controllers, it is easy to see that their web sites store much more information, as an example: the IP address, browser type, and device name. In the data collected by almost all controllers appear, aside from the information mentioned above, also the last session log-in with the related information about the IP, the access time, and the session time. 
However, looking better at the information we collected from most of the data controllers, they seem to be more a collection of obvious data they had than the result of an inquiry of personal data they collect. For example, in the case of e-shops, all the web sites return our transaction history, but only one returns the list of our research.
Once again, looking back at their privacy policy and thinking of the re-marketing techniques adopted, it is clear that they also have information about our research on their site.

Every data controller provides information with different granularity. For instance, if we compare two MOOC platforms (Massive Open Online Courses), one returns only the course to which the user is enrolled, the other returns specific information for each learning session.
Some giants of the internet have provided the most interesting information: who paid to send advertisement to the user, and what criteria have been used to target the user.

\subsubsection{Response readability}
As we said, almost half of the controllers provided the response as a structured file. Unfortunately, it is not easy to understand what kind of information is reported on it. Normally, in a structured data format each data is associated with one label. While understanding the meaning of some labels is quite intuitive, for most of the data provided in this way it is very hard to understand what kind of information the value represents even by users with a technical background. We found only 5 controllers that have attached to their structured response a file in which it is explained how to interpret the data.

\subsubsection{Right of data access vs right of data portability}
As we stated in~\ref{sec:resp_format}, in some cases, we requested both the data access and the data export. Comparing the two sets of data, we find that the data export has much more information. For instance, in the data access file only account information has been provided. Instead, in the file of the data export are present also details about every session we have done, the IP, and information related to our operating system and the browser. So, it is unclear why this information is provided only with one of the two requests.

\subsubsection{Third party tracker}
Nowadays, almost every web sites includes in the pages several scripts that collect information about the users or inject cookies that track them across the web. Among all the data collected, only 6  collectors provided such information. Even though these data is used to profile the users, and the web sites exploit that information to deliver targeted content, the web site itself is not the owner of the data. This kind of data decentralization means that the user who wants to know all the information that a web site handles about him has to perform a separate subject access request to all the services integrated into the web site. Unfortunately, this solution is almost impractical because of the large number of third-party services used by the web sites---more than 30 on average if we consider the private news media~\cite{sorensen2019before}. 

\subsubsection{Email tracking}
Among all the responses we obtained, only 2 data controllers provided us with a document containing information about the advertisement or communication emails they send and the information about email tracking~\cite{knox2010method}.
Through this technique it is possible for the sender to infer a large amount of private data. From a single email, it is possible to know information such as the IP and the device of the user. By sending with a proper timing the email, and analyzing the data, it is possible to infer the daily routine of the user, estimate the geographical area where he lives and where he works~\cite{xu2018privacy,englehardt2018never}.
So we believe that this information should be considered private data that the data controllers must provide to the owner only. 
To be sure that the email received was tracked and we interacted with them, we installed on our browser the Ugly Email\footnote{https://uglyemail.com/} tool. This tool scans the raw email looking for tracking elements. When it finds one, it raises an alert. Even though this tool is not able to detect every kind of tracking, no one of the data controllers we detected using this technology provides information about the email tracking.

\section{Privacy concerns}
In this section, we discuss the privacy concerns that emerged by our analysis of the procedures to provide the personal data and to identify the requesting user. We find that more than 50\% of data collectors that handled our request suffer from flaws that can compromise the users' privacy.

\subsection{Sharing data via email and no email encryption}
\label{subsec:mail}
We received most of the personal data by email. 82 of these shared the data as a plain file or a zip folder without using any security measure. Sending sensitive data as plain email or plain attachment can be risky, in fact, the email can be sent to an incorrect recipient. Moreover, since the email and the attached file are saved on the email server, there is the risk that an attacker gains unauthorized access to it, as it happened in the recent past~\cite{databreach}.
Sending personal data as an encrypted file is a best-practice encouraged by companies, universities~\cite{stanfordrisk}, government authorities~\cite{icoorg}, or as part of the GDPR interpretation~\cite{daninterpretation}.

Instead, among the data collectors that send the personal data encrypted via email, 20 of them send also the password to decrypt the data on the same email account, or even in the same email with the data attached. This solution is clearly ineffective.

We also found 3 interesting cases, where the controllers correctly encrypt the data and send the password on a different channel. However, a careful observer can quickly note that the passwords used to encrypt the data follow a pattern based on the requester data. Examples of these patterns are: user's surname concatenated to the same string, user's date of birth, or the user's full name. We double-check these patterns requesting the personal data to these controllers from 3 different accounts.

Finally, 2 of the collectors that use the email channel to provide the data, neither encrypt the file containing the personal data nor use TSL or s/MIME schemes to send the email, exposing the personal data to sniffing or a man-in-the-middle attacks.

\subsection{Identity card}
Most of data collectors base the access data request on the email exchange between the designed data protection officer and the user. To verify the identity of the users, we found that 33 of them required the scan of an ID as proof of identity. Among of them 2 required to send the scan of two different IDs.
Although this identification process could appear reasonable at first sight, it raises many concerns.
All the data collectors that ask for a proof of identity via email, do not provide a secure form or guidelines on how to send the ID, exposing the user to the threats described in section~\ref{subsec:mail}.
Moreover, it is unknown if the document sent is adequately stored to preserve the user privacy, deleted after the authentication phase, or if it remains on the email server.

During our investigation, an ID was also asked by data collectors that have no information about our identity, so it is unclear how the ID can help them to establish the ownership of the account.
Finally, we notice that for all those collectors that required a document, except one related to the cryptocurrency world, it is enough a photo of the ID's front page and does not provide any guidelines on how to take it.
This behavior raises the suspect that data collectors do not have tools to verify the authenticity of the document. Many IDs have the serial number on the back page. While through a photo it is not possible to verify most of the security features of the documents such as holograms, watermarks, or optically variable ink. 

\subsubsection{Tampering the document}
\begin{figure}
\begin{center}
\includegraphics[width=0.45\textwidth]{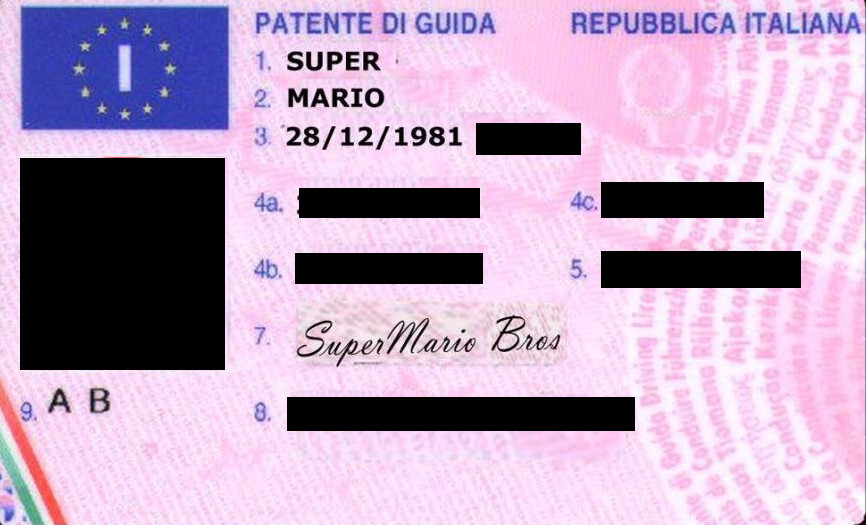}
\caption{A "Super Mario" example of obfuscated document. From the driving license we hide the owner's: photo, city of birth (field 3), date of release (field 4a), expiring date (field 4b), issuing entity (field 4c), ID number (field 5) residence address (field 8). }
\label{fig:obfuscated}
\end{center}
\end{figure}

\begin{figure}
\includegraphics[width=0.48\textwidth]{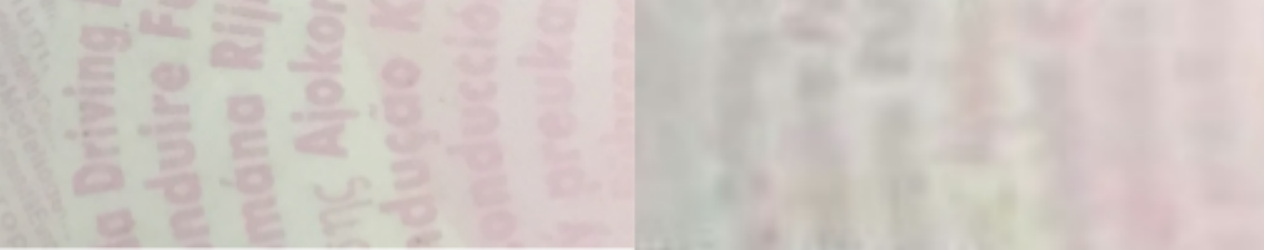}
\caption{A detail of the original photo of the identification document (on the left) and the same detail after we scaled down the quality of the image (on the right).}
\label{fig:id_details}
\end{figure}

Probably the best way to verify our intuition is to forge an identity document, but for ethical reasons, we do not choose this way. So, we altered a document of one of the authors in order to have an image of the document that can be considered even worse of a well forged one.
In particular, we manipulate our own identification ID in the following way:
\begin{enumerate}[label=(\alph*)]
    \item We pixelize the document, reducing the quality of the image scaling down its resolution. In particular, we substitute each square of 10x10 pixels with a single-pixel valued with the average of all the pixels contained in the mask. Fig~\ref{fig:id_details} shows a detail of an ID document before and after the scaling.
    \item We completely hide all the information contained in the document, even the owner's picture, putting on top of them a black layer. We leave only the full name and the date of birth of the document owner. Fig~\ref{fig:obfuscated} shows an example of an obfuscated document. Then we also scaled down the document, to have a pixelized image with obfuscated information.
\end{enumerate}
Once we obtain our tampered documents, we perform a SAR request to 25 out of 33 data controllers that required a document. We leave out from the experiment the data controllers that required 2 identification documents and the 6 that use a combination of identification procedures.
We send to all them a document of type (b), if they refuse the document because it is redacted, we carry a document of type (a).
At the end of the authentication phase, 21 out of 25 data controllers accepted our documents of type (b). While, the 4 data controllers that refused the document, accepted the ones of type (a), and also provide us the data. 
These results clearly show that most of the data controllers do not have proper tools/procedures or put enough effort into validating the identification documents. Moreover, we believe that this kind of identification procedure, even though it is powerful if appropriately applied, with the actual implementation it only exposes the users to useless risks.

\subsection{Email as authentication}
Nowadays, most of the online services are tied to the email account, so its access credentials should be kept as secret as the pin of a bank account. However, analyzing the data from data breaches it is easy to note that a large number of users use trivial passwords. As an example, 23.2 million victim accounts worldwide used '123456' as password~\cite{databreach}. At the same time, it is also well known that users tend to reuse the same password on different web services~\cite{wang2018next}, and so even if there is no breach from the email provider, a breach on another service can compromise the email account as well.

Along with our study, we found that 51 data controllers do not require additional information to the requester if the request comes from the same email address used to sign-up on the web site. This lack of control can be prone to spoofing attacks.
Since these 51 web sites seem to put less care in user identification, we want to understand their policy better.

To understand the robustness of their identification process, we simulate an attack on these data controllers. In our model, the attacker knows only the full name of the victim and the email address. So, we consider that the attacker achieves his goal to steal the private data of the victim only if the data controllers do not perform any kind of check about the identity of the requester.

Them, to perform the attack, we create a new email account, that looks very similar to the original one. As an example, the original one is "johndoe@provider.com" and the new one is "johnndoe@provider.com". Moreover, we set up the email account to display the name of the victim. 
Finally, in our request, we explicitly ask the data belonging to the original email address, and we sign all the emails with the name of the victim.

We contact all the 51 data controllers that apparently base their user identification on the email address, and we perform a SAR on behalf of the original account. 
Table~\ref{tab:email_attack} shown the number of data controllers we investigated, and the number of those vulnerable according to their category. 
As a result, we obtain that 19 data controllers provide us the data of the targeted account without carrying out other further identification processes. The others follow up the request with an identity verification phase. More in details 6 data controllers ask for an ID, 10 ask to perform the request from the original email address, 3 made knowledgeable questions, 5 send a confirmation email on the real email address and stop the communication, 2 requests to perform the request from inside the web site. 1 data controller that belongs to the adult category, follows up our malicious requests in an unorthodox way. Indeed, the DPO informed us about the kind of data stored by the web site, and notified to us that he deleted the victim's account.
To be sure that it wasn't a deceptive answer, as a response to our attack, we double-checked if the account had been actually deleted, and it had.
Notice that when we made the same request from the original email account, they provided us all the information requested. Now, even though this countermeasure is effectively against our attack, it is not fair towards the real owner of the account that lost the account without any communication.
Finally, the remaining 5 data controllers simply ignored the request.

With regards to the 19 data controllers that disclosed the personal data, we are sure that at least 10 of them noticed that the email address of the sender does not match with the one for whom we requested the data. In fact, 0 of these data controllers initially answered to our SAR that they have no data related to the fake email address. Then, after we pointed out to them that our request is on behalf of another email address, they share on the fake email address the personal data of victim.
The last data controller, in addition to sending us the data of the victim, he also updated the log-in credential with the fake email.
\begin{table}
	\centering
	\small
    \caption{%
      Fake email attack results.
    }\label{tab:email_attack}
    \begin{tabular}{l c c}
        \toprule
        Category & (\#)Attacked & (\#)Vulnerable \\
        \midrule
        Adult  & 4 & 0 \\
        Art & 4 & 1\\
        Business & 4 & 0 \\
        Computer & 2 & 0\\
        Health & 1 &1 \\ 
        Games & 4  & 3\\
        Kids \& Teen & 1 & 0\\
        News & 4  & 1\\
        Recreation & 5 & 3\\
        Reference & 3 & 3\\ 
        Science & 2 & 1\\
        Shopping & 11 & 2\\
        Society & 3 & 3\\
        Sports  & 3 & 1\\
        \bottomrule
        Total & 51 & 19 
    \end{tabular}
\end{table}
\subsection{Data escalation}
\label{sec:data_escalation}
In the light of the flaws we found, it is clear that an attacker, even with partial knowledge about the victim like full name and email, can jeopardize his privacy.
The attacker can build up a chain of requests to gather a lot of personal data. Most vulnerable data controllers could provide initial information such as the date of birth, interests, the IP address, and so the area where the user lives. From this information, the attacker can forge an obfuscated identity document. Doing so, the attacker can acquire additional personal data and refine the document. As the knowledge about the victim increases, the attacker can start to perform data requests even to the websites that use stronger identification procedures. Even if it is not possible to cheat all the data controllers, the total amount of possible information retrieved can still be remarkable. The feasibility of this kind of attack has been proved in~\cite{pavur2019}.

\section{Ethical Considerations}
In this work, we analyzed 334 data controllers that appear in the top position of the Alexa rank.
In all the experiments, we only tried to access data related to the authors of this work. We neither forge any identification document nor carried out an active attack against the server or the network of the targeted data controllers.
Consequently, and accordingly to the policy of our IRB, we did not need any explicit authorization to perform our experiments.

At the light of the concerning results we obtained, even though we neither disclose the entities affected by flaws nor we reported them to the authorities, we firmly believe that the noticed issues should be fixed for the privacy of the end-users.
So, we got in touch with the data controllers affected by flaws to perform a responsible disclosure. We sent to each data controller an email describing our research, the flaws we noticed, how to reproduce them, and suggesting countermeasures and best practices. 
 
\section{Related Work}

Since the GDPR entered into force one year ago, its effect on the web sites and companies has been studied from several prospective.
Sorensen et al.~\cite{sorensen2019before} analyze how the GDPR impacts on the presence of third-party trackers on web sites. They collected data belonging on 1250 popular web sites among a span range of four months before and after the GDPR entered into force. They observed that in the time range of observation, some categories such as private news, shopping travel, and entertainment presented a reduction in terms of the unique number of third-party web services, while in other categories, this number increases. Hence, they conclude that the GDPR does not clearly affect the presence of third-party on web sites. 

Wong et al.~\cite{wong2018portable} investigate the compliance of the files format after a request for data portability (Article 20) on 230 data controllers. Once achieved, all the responses they found that only 40\% of data format is compliant with the GDPR law, and 55\% of data controllers comply with the GDPR.

Urban et al.\cite{urban2019study} focused on the response time aspect of the GDPR. They perform a request to access to their personal data to 38 different tracking services. At the end of their investigation, they get that only 55\% of the companies targeted handled the request within the required time, while only 24\% provided the data.
In the Black Hat '19 event, Pavur et al.~\cite{pavur2019} show that it is possible to exploit the GDPR to jeopardize the privacy of the users. In particular, they perform SAR to 150 data controllers. They send to them a letter designed with the explicit intent to be vague and complex to satisfy, with the intent to distract the controller from the identity verification aspects of the law. They found that an attacker with few information about the victim, retrieved with open-source intelligence techniques, can get access to the data of 24\% of the data controllers under investigation. A similar attack was carried out in~\cite{di2019personal}. Here the author targeted 55 Belgian organizations. They were able to impersonate the victim and get their personal data from 15 out of 55 controllers using several techniques, among which document tampering and email spoofing. Finally, Cagnazzo et al.~\cite{cagnazzo2019gdpirated} exploiting a social engineering flaw, was able to retrieve personal data from 10 out of 14 German companies. They forged an email address that looks like one of the victims. From the forged email address, they contact the companies to update personal info about the victim. Finally, some days later, they perform the subject access request from the forged email.

Differently from these works, for the best of our knowledge, we are the first to conduct a comprehensive world-scale investigation. In our work, we deal with all the phases involved in the process, from the privacy policy pages to the response analysis, to end with the individuation of flaws of the data controllers.

\section{Conclusion}
In this work, we conducted a thorough investigation of the actual deployment of the GDPR law. We investigated about 341 data controllers worldwide, 334 of which are in the top rank for their category according to the Alexa rank. 
Our results that, to the best our knowledge, are unique in terms of scale provide meaningful insights on the handling of the subject access request by the controllers. 
In particular, we found that, for several reasons, it was impossible to obtain the data for 36.14\% of the targeted web sites. The 88.6\% of controllers that addressed the request responded within the GDPR time constraint, and needed an average of 16.4 days to fulfill the request. In many cases the data are provided in structured files that are hard to interpret, without any guidelines on the file structure from the controller. Lastly, we take into account how controllers transmit the data to the users and how they identify the requester. Surprisingly, we found that almost 50\% of the data controllers that handled the request are affected by flaws that can compromise the users' privacy.

\section{Future Work}
As future work, we intend to analyze the relationship between the way the data controllers handle the subject access request and their position in the Alexa rank or the categories they belong to.
We believe that there is a need for a more in-depth analysis of the completeness of the information provided. Since it is hard to estimate what information the web sites store, as assessment could be done comparing the information they declare to store in the privacy policy pages, with the ones retrieved with a subject access request.
In this work, we targeted web sites that are the most important in terms of web traffic. Moreover, most of them belong to big companies that have resources to update their systems to comply with the GDPR.  
As future work, it is interesting to investigate web sites that are at the lower positions of the rank. We believe that it is also appealing to conduct the same analysis on mobile applications. Indeed, the mobile application markets are abundant in applications developed by small companies or independent developers. We suppose that these subjects are more prone to be non-compliant with the GDPR.

\bibliographystyle{./bibliography/IEEEtran}
\bibliography{./biblio.bib}

\end{document}